\documentclass[twocolumn,showpacs,preprintnumbers,amsmath,amssymb,prl,aps,reprin,superscriptaddress]{revtex4-1}
\usepackage{graphicx}
\usepackage{amsmath}
\usepackage{verbatim}

\newcommand{\dB}{\mathrm{dB}}
\newcommand{\Ma}{M_0}
\newcommand{\Mb}{M_1}
\newcommand{\Mc}{M_2}
\newcommand{\QA}{\mathrm{Q}_{\mathrm{A}}}
\newcommand{\QB}{\mathrm{Q}_{\mathrm{B}}}
\newcommand{\wrf}{\omega_{\mathrm{RF}}}
\newcommand{\kHz}{\mathrm{kHz}}
\newcommand{\MHz}{\mathrm{MHz}}
\newcommand{\GHz}{\mathrm{GHz}}
\newcommand{\us}{\mu\mathrm{s}}
\newcommand{\ns}{\mathrm{ns}}
\newcommand{\ket}[1]{\left\lvert #1 \right\rangle}

\begin{document}

\title{Initialization by measurement of a two-qubit superconducting circuit}
\author{D. Rist\`e}
\author{J.~G. van Leeuwen}
\affiliation{Kavli Institute of Nanoscience, Delft University of Technology, P.O. Box 5046,
2600 GA Delft, The Netherlands}
\author {H.-S.~Ku}
\author {K.~W. Lehnert}
\affiliation{JILA, National Institute of Standards and Technology and the University of Colorado and Department of Physics, University of Colorado, Boulder, Colorado 80309, USA}
\author{L. DiCarlo}
\affiliation{Kavli Institute of Nanoscience, Delft University of Technology, P.O. Box 5046,
2600 GA Delft, The Netherlands}
\date{\today}

\begin{abstract}
We demonstrate initialization by joint measurement of two transmon qubits in 3D circuit quantum electrodynamics. Homodyne detection of cavity transmission
is enhanced by Josephson parametric amplification to discriminate  the two-qubit ground state from single-qubit excitations non-destructively  and with $98.1\%$ fidelity. Measurement and postselection of a steady-state mixture with $4.7\%$ residual excitation per qubit achieve $98.8\%$ fidelity to the ground state, thus outperforming passive initialization.
\end{abstract}

\maketitle
The abilities to initialize, coherently control and measure a multi-qubit register set the overall efficiency of a quantum algorithm~\cite{Nielsen00}.  In systems where qubit transition energies significantly exceed the thermal energy, initialization into the ground state can be  achieved by waiting several multiples of the qubit relaxation time $T_1$~\cite{Ladd10}. While this passive method has been standard in superconducting qubit systems,  recent breakthrough $T_1$ improvements~\cite{Paik11} in circuit quantum electrodynamics (cQED)~\cite{Blais04,Wallraff04} now bring its many shortcomings to light. First, commonly observed~\cite{Palacios-Laloy09,Geerlings12} residual qubit excitations can produce initialization errors exceeding the lowest single- and two-qubit  gate errors now achieved ($<0.3\%$~\cite{Paik11} and $<5\%$~\cite{Chow12}, respectively). Second, the wait time between computations grows proportionally with $T_1$.  Third, moving forward,  multiple rounds of quantum error correction~\cite{Schindler11} will require  re-initialization of ancilla qubits fast compared to coherence times.

Active means of initialization currently used in superconducting qubits include microwave sideband cooling~\cite{Valenzuela06,Manucharyan09}, temporal control~\cite{Reed10b} of Purcell-enhanced relaxation~\cite{Houck08} and coupling to spurious two-level systems~\cite{Mariantoni11}.
An attractive, but challenging alternative is to use a high-fidelity, quantum nondemolition (QND) readout~\cite{Braginsky96} to collapse qubits into known states. QND readout, already demonstrated for trapped ions~\cite{Hume07}, NV centers in diamond~\cite{Neumann10,*Robledo11}, and photons~\cite{Nogues99,*Johnson10,*Pryde04,*Johnson10}, also
opens the way to real-time quantum feedback~\cite{Wiseman09} and measurement-based quantum computing~\cite{Nielsen00}, and facilitates the  study of quantum jumps~\cite{Gleyzes07,Vijay11} and the Zeno effect~\cite{Gambetta08,Matsuzaki10}.
In cQED, significant progress in this direction has been achieved using bifurcation in nonlinear resonators~\cite{Mallet09} and parametric amplification~\cite{Vijay11,Abdo11}, but $T_1$ has until now limited the best QND readout fidelity to $86\%$.

In this Letter, we demonstrate ground-state initialization of two superconducting qubits by joint measurement and postselection. We combine long-lived transmon qubits in a 3D cQED architecture~\cite{Paik11} with phase-sensitive parametric amplification~\cite{Castellanos-Beltran08,Vijay09} to realize a high-fidelity, nondemolition readout. Homodyne measurement of cavity transmission at $\sim10$ intra-cavity photons discriminates the two-qubit ground state from single-qubit excitations with  $98.1\pm0.3\%$ fidelity (limited by $T_1$) and up to $99.6\%$ correlation between the measurement result and the post-measurement state.  We use this readout to purify the two-qubit system against a residual excitation of  $\sim4.7\%$ per qubit, achieving probabilistic ground-state preparation with $98.8\%$ fidelity.  During preparation of this manuscript, Johnson {\it et al.}~\cite{Johnson12} have reported similar results on initialization by measurement of one flux qubit in a 2D architecture.

Our system consists of an Al 3D cavity enclosing two superconducting transmon qubits, labeled $\QA$ and $\QB$, with transition frequencies $\omega_{\mathrm{A(B)}}/2\pi = 5.606~(5.327)~\GHz$, relaxation times $T_{1\mathrm{A(B)}}=23~(27)~\us$, and Ramsey dephasing times $T_{2\mathrm{A(B)}}^*=0.45~(4.2)~\us$~\cite{T2}. The fundamental mode of the cavity (TE101) resonates at $\omega_{r}/2\pi=6.548~\GHz$ (for qubits in ground state) with $\kappa/2\pi = 430~\kHz$ linewidth, and couples with $g/2\pi \sim 75~\MHz$ to both qubits. The measured dispersive shifts~\cite{Wallraff04}
$2\chi_\mathrm{A(B)}/2\pi=-3.7~(-2.6)~\MHz$ place the system in the strong dispersive regime of cQED~\cite{Schuster07}.

Qubit readout in cQED typically exploits dispersive interaction with the cavity. A readout pulse is applied at or near resonance with the cavity, and a coherent state builds up in the cavity with amplitude and phase encoding the multi-qubit state~\cite{Wallraff04,Majer07}. We optimize readout of $\QA$ by injecting a microwave pulse through the cavity at $\wrf=\omega_{r}-\chi_\mathrm{A}$, the average of  the resonance frequencies corresponding to qubits in $\ket{00}$ and $\ket{01}$, with left (right) index denoting the state of $\QB$ ($\QA$) [Figs.~1(a) and 1(d)]. This choice maximizes the phase difference  between the pointer coherent states. Homodyne detection of the output signal, itself proportional to the intra-cavity state, is overwhelmed by the noise added
by the semiconductor amplifier (HEMT), precluding high-fidelity single-shot readout [Fig.~1(c)]. We introduce a Josephson parametric amplifier (JPA)~\cite{Castellanos-Beltran08} at the front end of the amplification chain to boost the readout signal by exploiting the power-dependent phase of reflection at the JPA [see Figs.~1(a) and 1(b)]. Depending on the qubit state, the weak signal transmitted through the cavity is either added to or subtracted from a much stronger pump tone incident on the JPA, allowing single-shot discrimination between the two cases  [Fig.~1(c)].

\begin{figure}
\includegraphics[width=\columnwidth]{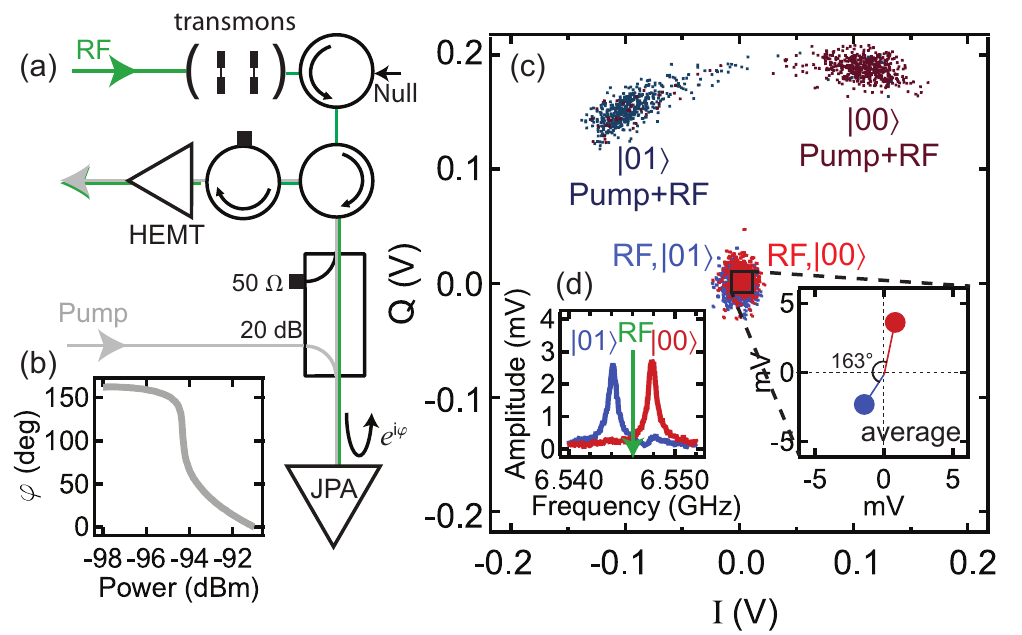}
\caption{(color online). JPA-backed dispersive transmon readout. (a) Simplified diagram of the experimental setup, showing the input path for the readout signal carrying the information on the qubit state (RF, green) and the stronger, degenerate tone (Pump, grey) biasing the JPA.  Both microwave tones are combined at the JPA and their sum is reflected with a phase dependent on the total power (b), amplifying the small signal. An additional tone (Null) is used to cancel any pump leakage into the cavity. The JPA is operated at the low-signal gain of $\sim25~\dB$ and $2~\MHz$ bandwidth. (c) Scatter plot in the $I-Q$ plane for sets of 500 single-shot measurements. Light red and blue: readout signal obtained with an RF tone probing the cavity for qubits in $\ket{00}$ and $\ket{01}$, respectively.
Dark red and blue: the Pump tone is added to the RF. (d) Spectroscopy of the cavity fundamental mode for qubits in $\ket{00}$ and $\ket{01}$. The RF frequency is chosen halfway between the two resonance peaks, giving the maximum phase contrast ($163^\circ$, see inset on the right).}
\end{figure}
The ability to better discern the qubit states with the JPA-backed readout is quantified by collecting statistics of single-shot measurements. The sequence used to benchmark the readout includes two measurement pulses, $\Ma$ and $\Mb$, each $700~\ns$ long, with a central integration window of $300~\ns$ [Fig.~2(a)].
Immediately before $\Mb$, a $\pi$ pulse is applied to $\QA$ in half of the cases, inverting the population of ground and excited state [Fig.~2(b)]. We observe a dominant peak for each prepared state, accompanied by a smaller one overlapping with the main peak of the other case.
We hypothesize that the main peak centered at positive voltage corresponds to state $\ket{00}$, and that the smaller peaks are due to residual qubit excitations, mixing the two distributions. To test this hypothesis, we first digitize the result of $\Ma$ with a threshold voltage $V_\mathrm{th}$, chosen to maximize the contrast between the cumulative histograms for the two prepared states [Fig.~2(c)], and assign the value $H (L)$ to the shots falling above (below) the threshold. Then we only keep the results of $\Mb$ corresponding to $\Ma=H$. Indeed, we observe that postselecting $91\%$ of the shots  reduces the overlaps from $\sim6$ to $2\%$ and from $\sim9$ to $1\%$ in the $H$ and $L$ regions, respectively~[Fig.~2(d)]. This substantiates the hypothesis of partial qubit excitation in the steady state, lifted by restricting to a subset of measurements where $\Ma$ declares the register to be in $\ket{00}$. Further evidence is obtained by observing that moving the threshold substantially decreases the fraction of postselected measurements without significantly improving the contrast [$\sim+0.1~ (0.2)\%$ keeping $85 ~(13)\%$ of the shots].
Postselection is effective in suppressing the residual excitation of any of the two qubits, since the $\ket{01}$ and $\ket{10}$ distributions are both highly separated from $\ket{00}$, and the probability that both qubits are excited is only $\sim0.2\%$~\cite{SOM}.
\begin{figure}
\includegraphics[width=\columnwidth]{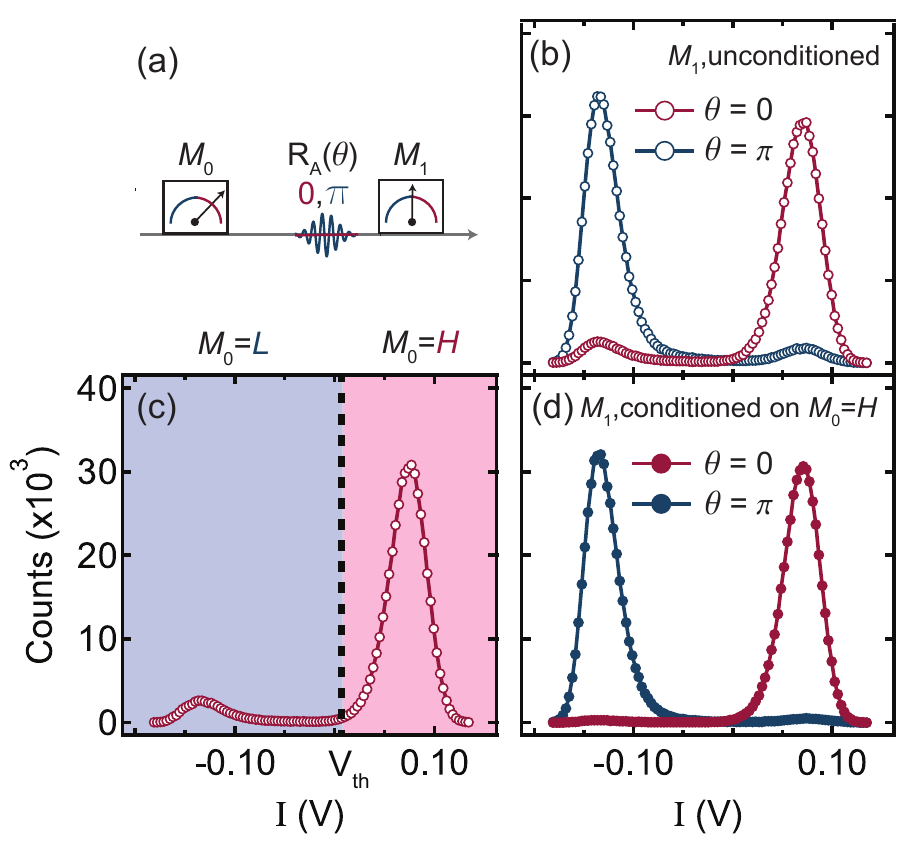}
\caption{(color online). Ground-state initialization by measurement. (a) Pulse sequence used to distinguish between the qubit states ($\Mb$), upon conditioning on the result of an initialization measurement $\Ma$. The sequence is repeated every $250~\us$. (b) Histograms of $500\,000$ shots of $\Mb$, without (red) and with (blue) inverting the population of $\QA$ with a $\pi$ pulse. (c) Histograms of $\Ma$, with $V_\mathrm{th}$ indicating the threshold voltage used to digitize the result. (d) $\Mb$ conditioned on $\Ma = H$ to initialize the system in the ground state, suppressing the residual steady-state excitation.
The conditioning threshold, selecting $91\%$ of the shots, matches the value for optimum discrimination of the state of $\QA$.}
\end{figure}

The performance of the JPA-backed readout and the effect of initialization by measurement are quantified by the optimum readout contrast. This contrast is  defined as the maximum difference between the cumulative probabilities for the two prepared states [Fig.~3(a)]. Without initialization, the use of the JPA gives an optimum contrast of $84.9\%$, a significant improvement over the $26\%$ obtained without the pump tone.
Comparing the deviations from unity contrast without and with initialization, we can extract the parameters for the error model shown in Fig.~3(b). The model (see the supplemental material), takes into account the residual steady-state excitation of both qubits, found to be $\sim4.7\%$ each, and the error probabilities for the qubits prepared in the four basis states. Although the projection into $\ket{00}$ occurs with $99.8\pm0.1\%$ fidelity, this probability is reduced to $98.8\%$ in the time $\tau=2.4~\us$ between $\Ma$ and $\Mb$, chosen to fully deplete the cavity of photons before the $\pi$ pulse preceding $\Mb$. We note that $\tau$ could be reduced by increasing $\kappa$ by at least a factor of two without compromising $T_{1\mathrm{A}}$  by the Purcell effect~\cite{Houck08}. By correcting for partial equlibration during $\tau$, we calculate an actual readout fidelity of $98.1\pm0.3\%$. The remaining infidelity is mainly attributed to qubit relaxation during the integration window.

As a test for readout fidelity, we performed single-shot measurements of a Rabi oscillation sequence applied to $\QA$, with variable amplitude of a resonant $32~\ns$ Gaussian pulse preceding $\Mb$, and using ground-state initialization as described above [Fig.~3(c)].
The density of discrete dots reflects the probability of measuring $H$ or $L$ depending on the prepared state. By averaging over $\sim10\,000$ shots, we recover the sinusoidal Rabi oscillations without (white) and with (black) ground-state initialization. As expected, the peak-to-peak amplitudes ($85.2$ and $96.7\%$, respectively) equal the optimum readout contrasts in Fig.~3(a), within statistical error.

\begin{figure}
\includegraphics[width=\columnwidth]{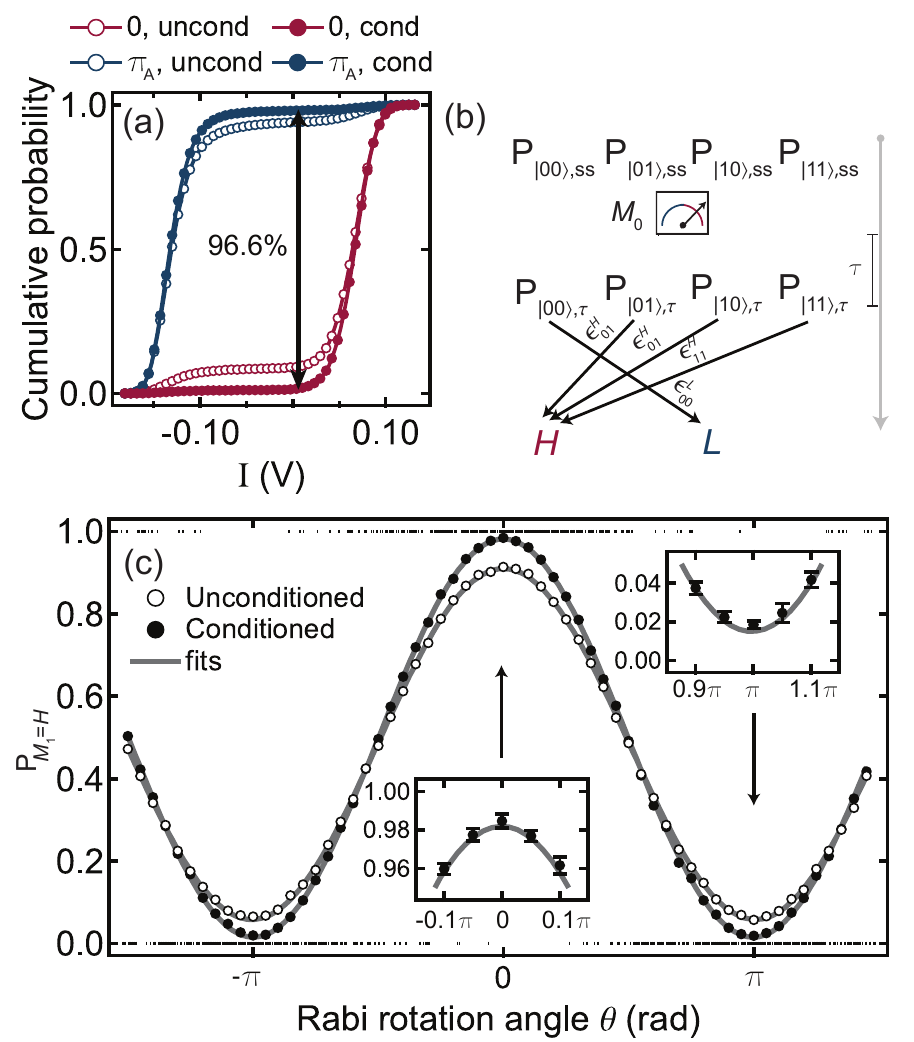}
\caption{(color online). Analysis of readout fidelity. (a) Cumulative histograms for $\Mb$ without and with conditioning on $\Ma=H$, obtained from data in Figs.~2(c) and 2(d). The optimum threshold maximizing the contrast between the two prepared states is the same in both cases. Deviations of the outcome from the intended prepared state are: $8.9\%$ ($1.3\%$) for the ground state, $6.2\%$ ($2.1\%$) for the excited state without (with) conditioning. Therefore, initialization by measurement and postselection increases the readout contrast from $84.9\%$ to $96.6\%$.
(b) Schematics of the readout error model, including the qubit populations in the steady state and at $\tau = 2.4~\us$ after $\Ma$. Only the arrows corresponding to readout errors are shown.
(c) Rabi oscillations of $\QA$ without (empty) and with (full dots) initialization by measurement and postselection. In each case, data are taken by first digitizing $10\,000$ single shots of $\Mb$ into $H$ or $L$, then averaging the results. Error bars on the average values are estimated from a subset of 175 measurements per point. For each angle, 7 randomly-chosen single-shot outcomes are also plotted (black dots at 0 or 1). The visibility of the averaged signal increases upon conditioning $\Mb$ on $\Ma=H$.}
\end{figure}

In an ideal projective measurement, there is a one-to-one relation between the outcome and the post-measurement state.
We perform repeated measurements to assess the QND nature of the readout, following Refs.~\onlinecite{Lupascu07,Boulant07}.
The correlation between two consecutive measurements, $\Mb$ and $\Mc$, is found to be independent of the initial state over a large range of Rabi rotation  angles $\theta$ [see Fig.~4(a)].  A decrease in the probabilities occurs when the chance to obtain a certain outcome on $\Mb$ is low (for instance to measure $\Mb=H$ for a state close to $\ket{01}$) and comparable to readout errors or to the partial recovery arising between $\Mb$ and $\Mc$. We extend the readout model of Fig.~3(b) to include the correlations between each outcome on $\Mb$ and the post-measurement state~\cite{SOM}. The deviation of the asymptotic levels from unity, $P_{H|H}=0.99$ and $P_{L|L}=0.89$, is largely due to recovery during $\tau$, as demonstrated in Fig.~4(b). From the model, we extrapolate the correlations for two adjacent measurements, $P_{H|H}(\tau=0)=0.996\pm0.001$ and  $P_{L|L}(\tau=0)=0.985\pm0.002$, corresponding to the probabilities that pre- and post-measurement state coincide. In the latter case, mismatches between the two outcomes are mainly due to qubit relaxation during $\Mc$.
Multiple  measurement pulses, as well as a long pulse, do not have a significant effect on the qubit state~\cite{SOM}, supporting the QND character of  the readout at the chosen power.
\begin{figure}
\includegraphics[width=\columnwidth]{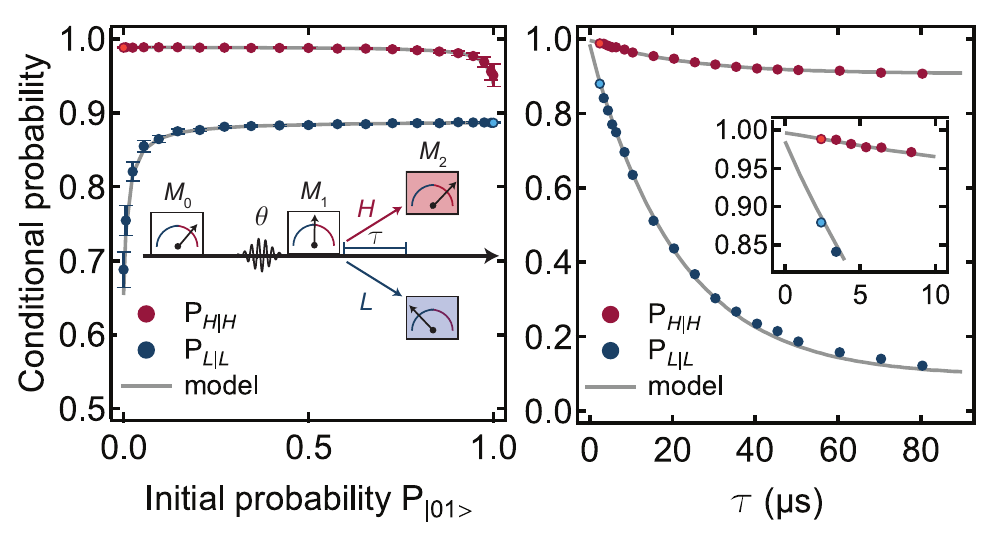}
\caption{(color online). Projectiveness of the measurement. (a) Conditional probabilities for two consecutive measurements $\Mb$ and $\Mc$, separated by $\tau=2.4~\us$. Following an initial measurement pulse $\Ma$ used for initialization into $\ket{00}$ by the method described, a Rabi pulse with variable amplitude rotates $\QA$ by an angle $\theta$ along the $x$-axis of the Bloch sphere, preparing a state with $P_{\ket{01}} = \sin^2(\theta/2)$. Red (blue): probability to measure $\Mc=H (L)$ conditioned on having obtained the same result in $\Mb$, as a function of the initial excitation  of $\QA$. Error bars are the standard error obtained from 40 repetitions of the experiment, each one having a minimum of 250 postselected shots per point. Deviations from an ideal projective measurement are due to the finite readout fidelity, and to partial recovery after $\Mb$~\cite{SOM}. The latter effect is shown in (b), where the conditional probabilities converge to the unconditioned values,  $ P_{H}=0.91$ and $ P_{L}=0.09$ for $\tau\gg T_1$, in agreement with Fig.~2, taking into account relaxation between the $\pi$ pulse and $\Mc$. Error bars are smaller than the dot size.}
\end{figure}

We have demonstrated the simultaneous projection by measurement of two qubits into the ground state. This technique allows us to correct for residual single-qubit excitations, preparing the register in $\ket{00}$ with $98.8\%$ probability. Initialization will be imperfect when the population of the doubly-excited state is relevant, a problem that can be addressed by choosing a different configuration of the joint readout, fully discriminating one of the computational states from the other three. A straight-forward extension of this work will use the knowledge gained by projection to condition further coherent operations on one or more qubits. 
For example, measuring a qubit and applying a $\pi$ pulse conditioned on having projected onto the excited state will deterministically prepare the ground state on a timescale much shorter than $T_1$.
Future experiments will also target the generation of entanglement by multi-qubit parity measurement~\cite{Lalumiere10,Tornberg10}.

\begin{acknowledgments}
We thank F.~Nguyen for discussions and experimental assistance, P.~C.~de Groot and M.~Shakori for fabrication support, and R.~N.~Schouten and G.~de Lange for electronics support. We  acknowledge funding from the Dutch Organization for Fundamental Research on Matter (FOM), the Netherlands Organization for Scientific Research (NWO, VIDI grant~680-47-508), the EU FP7 project SOLID, and the DARPA QuEST program.
\end{acknowledgments}


\begin{thebibliography}{10}
\bibitem{T2}
$\QA$ is a double-junction qubit with a random, but fixed magnetic flux offset
  placing it $\sim1~\GHz$ below from its flux sweet spot, limiting its $T_2^*$
  \cite{Schreier08}.
\end{thebibliography}
\end{document}


\title{Supplement to "Initialization by measurement of a two-qubit superconducting circuit"}
\author{D. Rist\`e}\
\author{J.~G. van Leeuwen}
\affiliation{Kavli Institute of Nanoscience, Delft University of Technology, P.O. Box 5046,
2600 GA Delft, The Netherlands}
\author {H.-S.~Ku}
\author {K.~W. Lehnert}
\affiliation{JILA, National Institute of Standards and Technology and the University of Colorado and Department of Physics, University of Colorado, Boulder, Colorado 80309, USA}
\author{L. DiCarlo}
\affiliation{Kavli Institute of Nanoscience, Delft University of Technology, P.O. Box 5046,
2600 GA Delft, The Netherlands}
\date{\today}

\maketitle

\section{Device fabrication}
The qubits are patterned on a sapphire substrate (C-plane, $430~\um$ thickness, single-side polished) by electron-beam lithography and Al double-angle evaporation with intermediate oxidation ($\mathrm{O}_2$, 10 min, $0.4~\mathrm{mBar}$). The qubit design is very similar to that pioneered in Ref.~\onlinecite{Paik11}, with junction sizes $\sim~300 \times 300~\nm^2$ and $250\times 500~\um^2$ antennas.  Qubit $\QA$  is a double-junction transmon, and qubit $\QB$ is of the single-junction type.
The 3D cavity is machined from Al alloy 6082 ($\mathrm{AlSi1MgMn}$) in two halves with internal cross section $35.4 \times 10.4~\mm^2$ and total height $24.3~\mm$.
\begin{figure}
\includegraphics[width=\columnwidth]{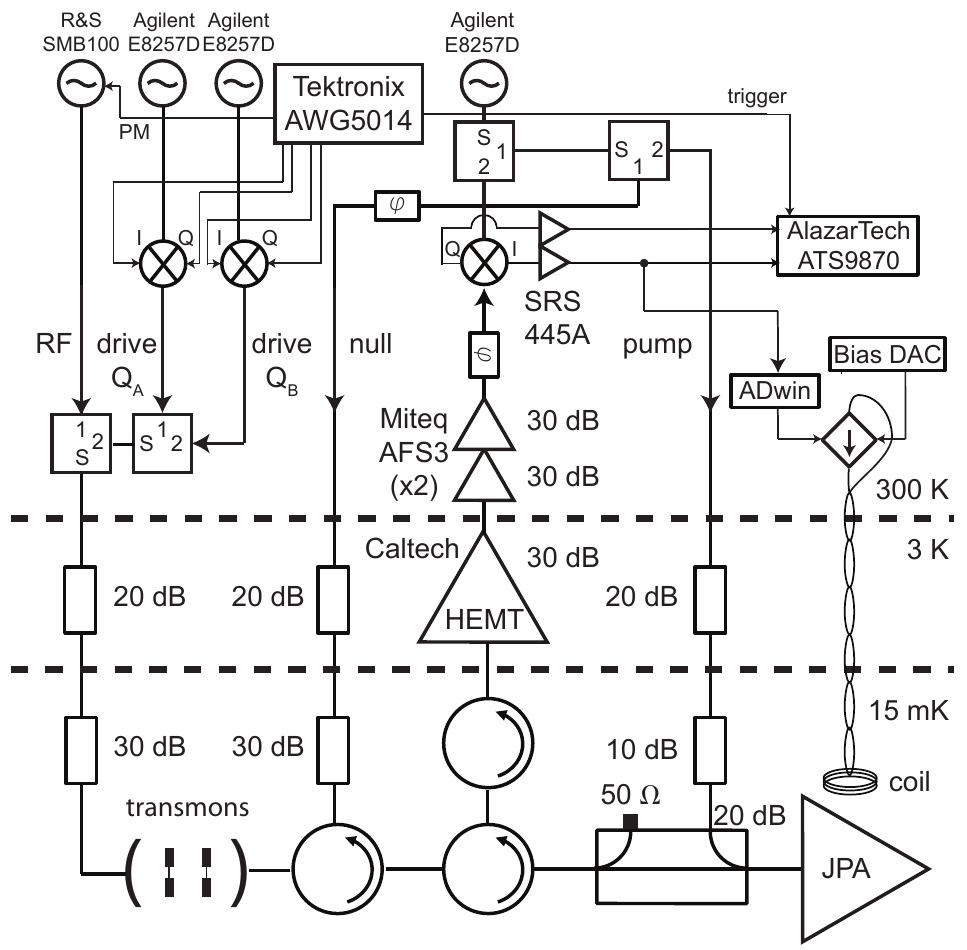}
\caption{Detailed schematic of the experimental setup. The measurement tone (RF) and qubit drives enter the cavity through the same transmission line. The RF is combined with a stronger pump tone ($\sim+26~\dB$) at the JPA, which is flux-biased with an external superconducting coil. The pump signal is split into two arms, one being directed to the cavity through a circulator, in order to suppress the photon leakage from the pump port. The readout signal is amplified at $3~\K$ (Caltech Cryo1-12, $0.06~\dB$ noise figure) and at room temperature (two Miteq amplifiers, $4-8~\GHz, 0.8~\dB$ and $2~\dB$ noise figure at $6.5~\GHz$. It is then demodulated ($0~\Hz$ intermediate frequency) and re-amplified (SRS445A, $25~\V/\V$ gain) before being digitized by an AlazarTech ATS9870 (1~GS/s, 8 bits). The ouptut signal is also fed to an ADwin-GOLD processor, running a proportional-integral control loop which actively stabilizes the JPA bias point.}
\end{figure}

\begin{figure}
\includegraphics[width=0.75\columnwidth]{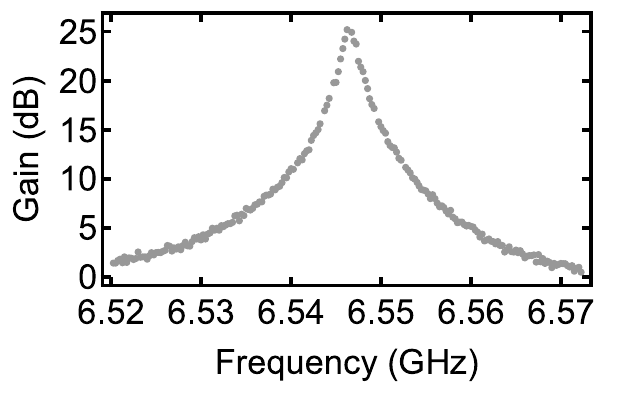}
\caption{Small-signal gain of the JPA at the bias point. The peak gain is $25~\dB$ and the full width at half maximum is $2~\MHz$.}
\end{figure}

\section{Calibration of the JPA for qubit readout}
The resonance frequency of the JPA is set by flux biasing its 20 SQUID loops ($25~\um^2$ flux-threading area in each)~\cite{Castellanos-Beltran08} with a home-built superconducting coil driven by a voltage-controlled current source. The bias point is actively stabilized with an ADwin-GOLD processor running  a proportional-integral feedback loop. This active control of the bias point is effective because  the duty cycle of measurement, which shifts the output level
[Figs.~S2 and 1(c)], is only $0.3\%$.

The JPA is operated in the phase-sensitive mode, with a continuous wave (CW) pump tone at the readout frequency, $\wrf/2\pi=6.5446~\GHz$ [Figs.~S1 and 1(c)]. The pump power ($-94~\dBm$) bends the JPA resonance lineshape from $6.564~\GHz$ down to $\wrf$, making the phase of the reflected signal highly sensitive to variations in the incident power [Fig.~1(b)]. A phase shifter in the output line is used to maximize the sensitivity of the in-phase (I) quadrature after demodulation. The relative phase between the measurement pulse and the pump is adjusted to maximize the contrast between the histograms for qubits prepared in $\ket{00}$ and $\ket{01}$. This calibration is repeated every $\sim5\min$ to cancel any phase drift between the two generators.
An additional CW tone (null), split from the pump generator, is injected to the cavity via the output port. This tone is used to cancel the leakage from the pump into the cavity arising from the limited isolation ($\sim40~\dB$ in total) of the two circulators between JPA and cavity. Cancellation is achieved by tuning the amplitude and phase of the null tone to suppress the residual one-photon peak visible in qubit spectroscopy~\cite{Schuster07}. The nulling is crucial to realize high-fidelity single-qubit pulses by avoiding Stark shifts. Nulling also prevents leakage photons from continuously measuring the qubits. The chosen power of the RF pulse corresponds to $\sim10$ intra-cavity photons in the steady state, calibrated by qubit spectroscopy (not shown), and is the minimum power that achieves a complete separation of the histograms at the chosen JPA bias point.

\begin{figure}
\includegraphics[width=0.85\columnwidth]{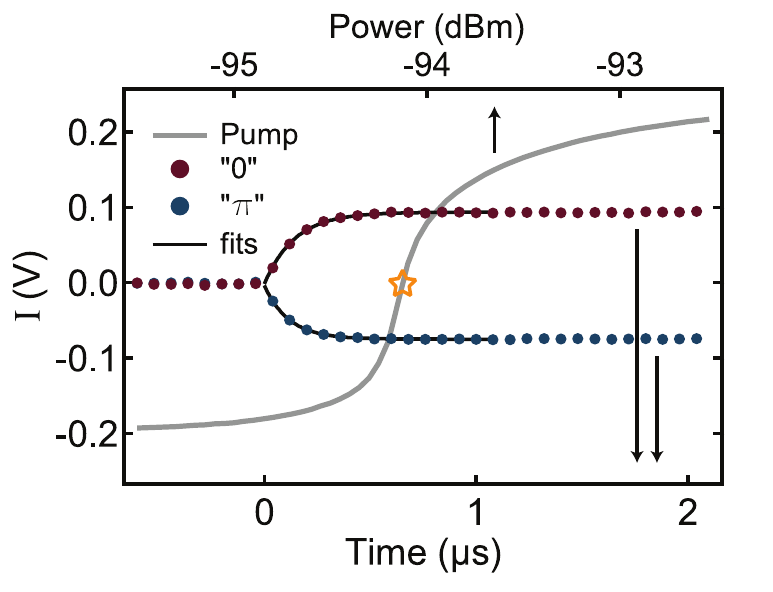}
\caption{Bias point and time response of the JPA. Grey (top axis): in-phase quadrature of the pump output as a function of incident power. The quadrature phase is chosen to have the maximum sensitivity to power at the bias point (orange star). Red and blue (bottom axis): average output signal response to a pulse added at the pump port. The RF signals corresponding to qubit states $\ket{0}$ and $\ket{1}$ are simulated by injecting two pulses with opposite phase. The power of the pulsed tones is chosen to match the contrast obtained in the same way when pulsing through the cavity. The measured response time of the JPA in this regime is $\sim130~\ns$ (average of the two time constants obtained from the fits), much shorter than $\kappa^{-1}=0.75~\us$ and $T_{1,\mathrm{A(B)}}=23 (27)~\us$.}
\end{figure}

\section{Readout error model}
We first characterize the readout errors by considering the probabilities of measuring $H$ or $L$ for the four input states. As shown in Figs.~2 and 3, the measured deviations from unity contrast are due to the combination of readout errors and steady-state excitations. From Fig.~3(a) we extract four values at the optimum readout threshold, one per prepared state, with and without initialization. We denote by $\epsilon^M_{ij}$ the probability of obtaining the measurement result $M$ ($H$ or $L$) for the pre-measurement state $\ket{ij}$, and by $P_{\ket{ij},\mathrm{ss}}$ and $P_{\ket{ij},\mathrm{\tau}}$ the populations of the qubits in $\ket{ij}$ in the steady state and at $\tau=2.4~\us$ following initialization, respectively. By using the measured  $T_{1\mathrm{A},\mathrm{B}}$ and taking $P_{\ket{ij},\mathrm{ss}}$ and $\epsilon^M_{ij}$ as free parameters, we calculate the two-qubit populations immediately before $\Mb$, with or without initialization, and the probabilities for each measurement outcome. Good agreement of the simulated sequence with the data is obtained with the values in the top half of Table~S1, also supported  by the experiment in Fig.~4(a) (see below).

\begin{figure}
\includegraphics{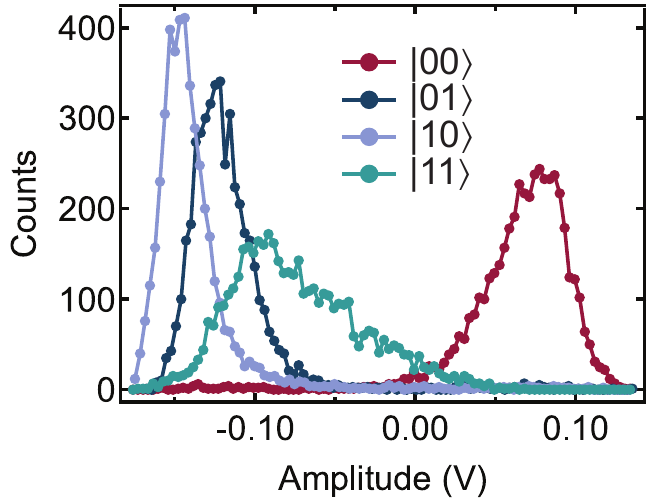}
\caption{Histograms of measurements following preparation of the two-qubit register in the four basis states. The states are prepared by applying $\pi$ pulses $2~\us$ after an initial measurement pulse used for initialization by postselection.  Each histogram contains $5\,000$ measurement shots.  The poor separation of $\ket{11}$ from $\ket{00}$ is due to the lower cavity transmission at $\omega_r - \chi_A - \chi_B$, giving a reduced shift of the pump from the bias point.}
\end{figure}
\begin{figure}
\includegraphics{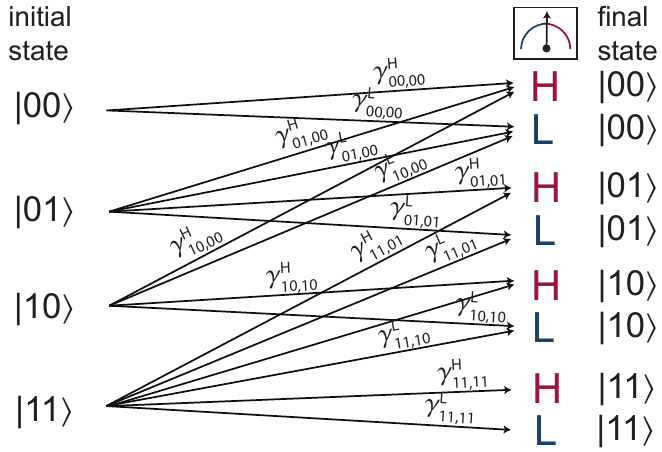}
\caption{The full readout error model. This model extends the one in Fig.~3(b) by including the post-measurement state. $\gamma^M_{ij,kl}$ indicates the probability of obtaining the post-measurement state $\ket{kl}$, given the initial state $\ket{ij}$ and the measurement result $M$ ($H$ or $L$).} 
\end{figure}
The readout process is further analyzed by considering the probabilities of measuring $H$ or $L$ for the four input states, and for all the possible post-measurement states (Fig.~S5). For example, $\gamma^H_{01,00}$ indicates the probability that measuring $\ket{01}$ gives the result $H$ and the post-measurement state is $\ket{00}$.
Again, we assume symmetry between the two qubits, i.e., $\gamma^M_{ij,kl}=\gamma^M_{ji,lk}$. Recovery to the steady state [Fig.~4(b)] and measurement-induced excitation (Fig.~S7) during the integration window are negligible, so we ignore the transitions from ground to excited state of any qubit during the measurement. Using the values extracted above and with  $\gamma^M_{ij,kl}$ as free parameters, we calculate the population of the four basis states at any point in the sequence of Fig.~4(a), including errors in the conditioning step $\Mb$, partial recovery to the steady state between $\Ma$ and $\Mb$ and between $\Mb$ and $\Mc$, and relaxation during measurements. The best agreement to the data (solid curves) is obtained with the values in the bottom half of Table~S1. Summing over the probabilities corresponding to different final states, we obtain the error parameters $\epsilon^M_{ij}$ = $\sum_{k,l} \gamma^M_{ij,kl}$, consistent with the values derived from Fig.~3(a). Bounds on the extracted parameters are defined by the confidence interval in which the model matches the data within the error bars. The high sensitivity of the model curve to the error parameters allows us to further reduce the uncertainty on the top half of the table as well. The effect of the probabilities involving both qubit excitations on the calculated populations, however, is marginal, so we can only estimate $\epsilon^H_{11}$ and $\gamma^H_{11,11}$.

\begin{table}
\caption{Extracted parameters of the readout model.}
\begin{ruledtabular}
\begin{tabular}{llll}
$P_{\ket{00},\mathrm{ss}}$ & $0.908\pm0.006$ & $P_{\ket{00},\tau}$ & $0.993\pm 0.006$\\
$P_{\ket{01},\mathrm{ss}},P_{\ket{10},\mathrm{ss}}$ & $0.047\pm0.003$ & $P_{\ket{01},\tau},P_{\ket{10},\tau}$ & $0.003\pm 0.001$\\
$P_{\ket{11},\mathrm{ss}}$ & $0.002\pm0.001$ & $P_{\ket{11},\tau}$ & $0.002\pm0.001$\\
$\epsilon^L_{00}$ & $0.004\pm0.001$ & $\epsilon^H_{01},\epsilon^H_{10}$ & $0.015\pm0.002$\\
$\epsilon^H_{11}$ & $0.15\pm0.15$ & &\\
\hline
$\gamma^H_{00,00}$& $0.996\pm0.001 $ & $\gamma^L_{00,00}$ & $0.004\pm0.001$ \\
$\gamma^H_{01,00},\gamma^H_{10,00}$& $0.015\pm0.005 $ & $\gamma^L_{01,00},\gamma^L_{10,00}$ & $0\pm0.001$ \\
$\gamma^H_{01,01},\gamma^H_{10,10}$& $0\pm0.001 $ & $\gamma^L_{01,01},\gamma^L_{10,10}$ & $0.985\pm0.002$ \\
$\gamma^H_{11,01},\gamma^H_{11,10}$& $0.02\pm0.02$ & $\gamma^H_{11,11}$& $0.11\pm0.05$ \\
\end{tabular}
\end{ruledtabular}
\end{table}

Further evidence of the QND character of readout is obtained by performing  consecutive measurements. The cavity response for qubits in the steady state or after a $\pi$ pulse on  $\QA$ shows no significant dependence on the number of pulses preceding a given measurement, indicating that the readout is nondestructive [Fig.~S6(a)]. Similarly, the cumulative histograms for the same cases exhibit substantial overlap in Fig.~S6(b), where the small difference is partly explained by the incomplete damping of the cavity between pulses. In the extreme case of a $70~\us$ readout pulse (Fig.~S7), much longer than the pulse length used in the experiment  ($700~\ns$), the measurement-induced qubit excitation is only $1\%$. This fraction increases drastically to $22\%$ doubling the readout power.

\begin{figure}
\includegraphics[width=1\columnwidth]{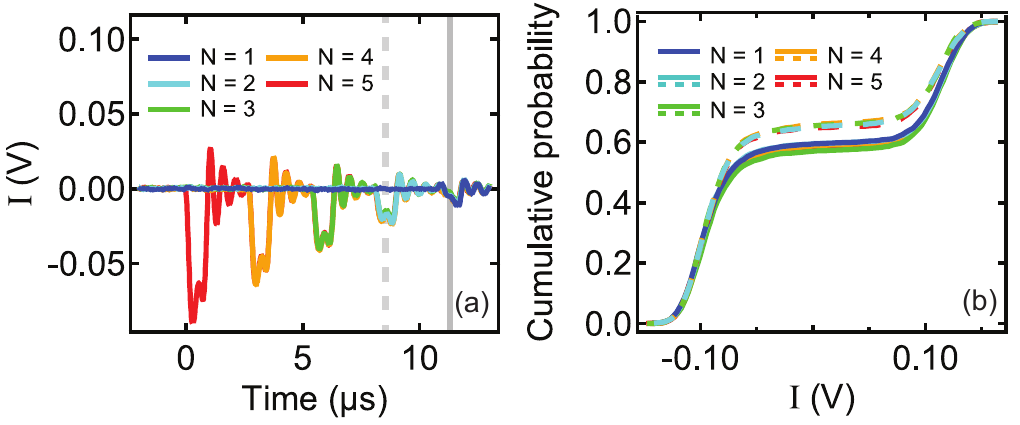}
\caption{Repeated measurements. (a) Averaged readout response ($20\,000$ repetitions) with the population of $\QA$ inverted at $t=0_{-}$ and $N\in [1,5]$ consecutive measurement pulses, each $700~\ns$ long and $2~\us$ apart from the next one. (b) Cumulative histogram of single shots for the same range of $N$, integrated over the two windows indicated by vertical lines in (a).}
\end{figure}

\begin{figure}
\includegraphics[width=0.85\columnwidth]{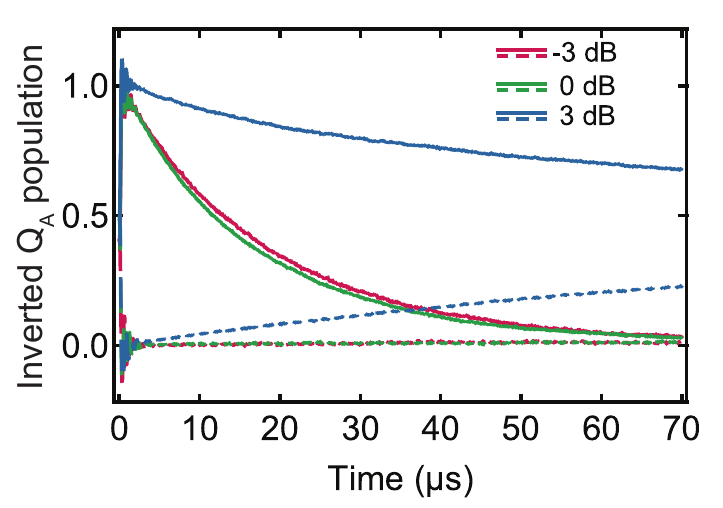}
\caption{Long measurements. Averaged readout response  ($20\,000$ repetitions) for a long measurement turned on at $t=0$, with qubits in the steady state (dashed curves) or following a $\pi$ pulse on $\QA$ (solid). Each pair of responses is normalized with the initial values with and without $\pi$ pulse. The indicated RF power is relative to that used in other data shown. At this power, the excited state decays in $17~\us$, while the ground state is excited only $1\%$ at $70~\us$, already $100$ times longer than the integration window. With $3~\dB$ less, the decay time is $18~\us$, and excitation is $1\%$. With $3~\dB$ more, the equilibration time constant after inversion is $39~\us$, and the excitation from steady state is $22\%$ at $70~\us$.}
\end{figure}